\documentclass[preprint,aps]{revtex4}

\usepackage{dcolumn}
\usepackage{amsmath}

\usepackage{graphicx}
\usepackage{bm}

\def\eref#1{(\ref{#1})}
\def\d{{\rm d}}
\def\e{{\rm e}}
\def\O{{\rm O}}

\begin{document}

\title{Dynamics of a simple evolutionary process}
\author{Dietrich Stauffer}
\affiliation{Institute for Theoretical Physics, Cologne University,
D-50923 K\"oln, Germany}
\author{M. E. J. Newman}
\affiliation{Santa Fe Institute, 1399 Hyde Park Road, Santa Fe, NM 87501,
U.S.A.}
\date{31 July 2001}

\begin{abstract}
We study the simple evolutionary process in which we repeatedly find the
least fit agent in a population of agents and give it a new fitness which
is chosen independently at random from a specified distribution.  We show
that many of the average properties of this process can be calculated
exactly using analytic methods.  In particular we find the distribution of
fitnesses at arbitrary time, and the distribution of the lengths of runs of
hits on the same agent, the latter being found to follow a power law with
exponent~$-1$, similar to the distribution of times between evolutionary
events in the Bak--Sneppen model and models based on the so-called record
dynamics.  We confirm our analytic results with extensive numerical
simulations.
\end{abstract}

%\pacs{Valid PACS appear here}
%\keywords{Suggested keywords}

\maketitle

\section{Introduction}
Is there any progress in the way humans live together?  Yee~\cite{Yee01}
has recently suggested a simple model for the development of
law-by-precedent, which assumes that bad decisions of legal courts have a
higher tendency to be overruled by later and/or higher court decisions,
thus improving the quality of the law over time.  Yee's model can be
thought of in general terms as an evolutionary process in which the least
fit individuals in a population are successively removed and replaced by
others~\cite{BS93}.  In this paper we study this process analytically,
providing some exact results and confirming these results with numerical
simulations.

In the model proposed by Yee, a large number of agents $i=1\ldots N$,
representing biological species, court decisions, firms, etc.~\cite{CVV01},
each possess a fitness or quality $e_i\in[0,1)$.  Initially the $e_i$ are
uniformly distributed over their range.  At each time-step, we find the
individual that has the lowest fitness in the population and give it a new
fitness which is chosen uniformly at random in the interval from zero to
one.  This model can be regarded as a simplified version of the
Bak--Sneppen model of coevolution~\cite{BS93,RJ94}, in which the least fit
agent in a population and its neighbours on a $d$-dimensional lattice or
other network are repeatedly removed and replaced with new agents with
randomly chosen fitnesses.  In the Bak--Sneppen model the agents typically
represent species of organisms, and the replacement of neighbours arises as
a result of interactions between species: host-parasite or prey-predator
interactions, for instance~\cite{SJ00}.  In the case of court decisions and
some other systems there is no strong case for such interactions and Yee
thus omitted them, replacing instead only the element with the lowest
fitness.  Amongst other things this means that Yee's model does not show
the self-organized criticality that is the principal focus of study in the
Bak--Sneppen model, but Yee's model still has non-trivial behaviour.

Yee's model is also reminiscent of the so-called ``record
dynamics''~\cite{Feller67,SL93}, which is the process which describes the
pattern formed by the highest value seen so far in a sequence of random
numbers.  In both models, the interesting behaviour comes largely from the
non-equilibrium nature of the dynamics.

It is worth noting that one does not need to choose fitnesses from the
uniform distribution for the results given in this paper to be valid.
Since the dynamics of Yee's model is determined solely by the ranking of
the agents relative to one another and is unaffected by their absolute
fitness, one is free to chose numbers from any (normalizable) distribution
and the results will be identical.  We use uniform random numbers solely
for ease of analysis and simulation.

\section{Analytic results}
The basic form that the evolution of Yee's model takes is clear.  Suppose
that $x(t)$ is the value of the lowest of the original distribution of
fitnesses $e_i$ which has not yet been touched at time~$t$.  Then the
distribution of fitnesses above this value must be uniform, since all
fitnesses are chosen at random from the uniform distribution.  At
time-step~$t$, finding the lowest fitness in the population, we replace it
with a random number which with probability $1-x$ falls above $x$ and hence
is not the new lowest fitness.  When this happens, the value of $x$
increases by an amount which is given by $(1-x)/N$ on average, and hence
the rate at which $x$ increases is
\begin{equation}
{\d x\over\d t} = {(1-x)^2\over N}.
\end{equation}
Defining a reduced time $\tau=t/N$, this has the solution
\begin{equation}
x(\tau) = {\tau\over\tau+1}.
\label{solnx}
\end{equation}
Thus, as we would expect, the lowest fitness in the population increases
monotonically and tends to~1 as $t\to\infty$.

This however does not tell the whole story.  As the value of $x$ increases,
the chances of a new randomly chosen fitness falling above $x$ decreases,
and so at longer times it takes more attempts at replacing the fitness
value $e_i$ of an agent to find one which falls above~$x$.  Thus the
dynamics of the model consists of ``runs'' of attempts at finding a new
higher value for the fitness of the least fit agent.

Let us define a ``run'' to be the number of consecutive attempts to improve
the fitness of a given agent until a new fitness is chosen which makes that
agent no longer the least fit individual in the population.  The
probability that a new run starts at time $t$ is equal to the probability
that the random number chosen at time $t-1$ fell above $x$, which is
$1-x=1/(\tau+1)$.  Thus the average length of a run at time $t$ is
$1/(1-x)=\tau+1$.  The total number of runs from time $0$ until a final
time $t_f=N\tau_f$ is
\begin{equation}
R = \sum_{t=0}^{t_f-1} {1\over t/N+1} = N\int_0^{\tau_f} {\d\tau\over\tau+1}
  = N \log(\tau_f+1),
\label{mean}
\end{equation}
where the integral becomes exact in the limit of large system size with
$\tau_f\gg1$.  Since each agent is equally likely to be the least fit on
each run, the average number of runs which affect each agent is
$R/N=\log(\tau_f+1)$.

We can also calculate the complete distribution of the number $k$ of runs
which affect any given agent.  Since, again, each agent is equally likely
to be the least fit on each run, this is simply a binomial distribution
\begin{eqnarray}
p_k &=& {R\choose k} \left[{1\over N}\right]^k
        \left[1-{1\over N}\right]^{R-k}
     \simeq  {R\choose k} {1\over N^k(\tau_f+1)}\nonumber\\
    &\simeq& {[\log(\tau_f+1)]^k\over k!(\tau_f+1)},
\label{binomial}
\end{eqnarray}
where we have made use of the value of $R$ from Eq.~\eref{mean}, and the
last two equalities become exact for large~$N$.  When $k$ is large, which
it normally will be, one can further use Stirling's approximation
$k!\simeq\sqrt{2\pi/k}\,(k/\e)^k$ to write this as
\begin{equation}
p_k \simeq {\sqrt{k/2\pi}\over\tau_f+1}
           \left[{\e\log(\tau_f+1)\over k}\right]^k.
\end{equation}

The distribution of the lengths of runs can also be calculated exactly.  At
time $t$, the lowest fitness so far untouched is $x$, and the probability
distribution of the lengths of runs is
\begin{equation}
p_n = x^{n-1} (1-x),
\end{equation}
which is correctly normalized, as can easily be verified.  Since the
probability that a new run starts at time $t$ is $1-x$, the overall
probability distribution of the lengths of runs for the entire lifetime of
the model is
\begin{equation}
P_n = \sum_{t=0}^\infty x^{n-1} (1-x)^2.
\end{equation}
Changing variables to $\tau$ again, using Eq.~\eref{solnx}, and taking the
limit of large $N$, the sum becomes an integral, and we find
\begin{equation}
P_n \propto \int_0^\infty {\tau^{n-1}\over(\tau+1)^{n+1}}\>\d\tau
    = {1\over n}.
\label{solnpn}
\end{equation}
The lengths of the individual intervals between the evolution of one agent
and the next have a non-stationary, monotonically increasing average as
time passes.  But Eq.~\eref{solnpn} indicates that if one calculates an
average over all times of the distribution of intervals, it follows a
power-law with exponent~$-1$.  This is reminiscent of the behaviour of the
record dynamics, which also shows such an $n^{-1}$ law when averaged over
all times~\cite{NFST97}.  An $n^{-1}$ law also appears in thermal
barrier-crossing processes, such as those governing the times between
evolutionary events in the Bak--Sneppen model.  And similar power-law
behaviour is seen in real-world macroevolution, where the distribution of
the times between evolutionary events is clearly
non-stationary~\cite{SSA95}.  If we equate evolutionary events with species
extinction, then the logarithmic behaviour~\eref{mean} in the cumulated
number of events and the power-law distribution of the intervals between
them~\eref{solnpn} seen in the model are identical to those seen in the
fossil record of extinctions~\cite{NE99}.

Note that the distribution~\eref{solnpn} is not integrable and therefore
cannot be normalized.  In any real case however---in simulations of the
model for example---the distribution is truncated by finite-time effects
and is perfectly normalizable.

\begin{figure}[t]
\begin{center}
\resizebox{4in}{!}{\includegraphics{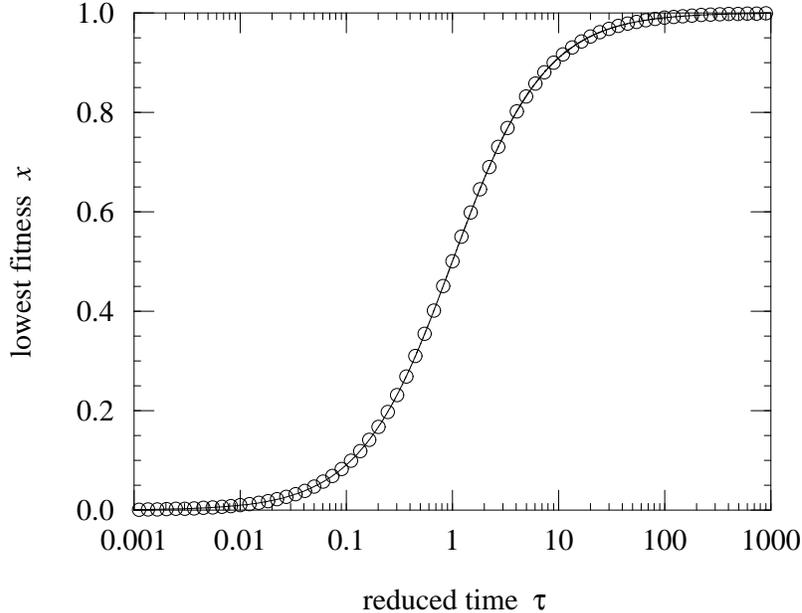}}
\end{center}
\caption{Fitness $x$ of the least fit member of the population as a
function of the reduced time $\tau=t/N$.  The points are simulation results
for $10^9$ time-steps with $N=10^6$, and the solid curve is the exact
solution in the large system-size limit, Eq.~\eref{solnx}.}
\label{xt}
\end{figure}

\begin{figure}[t]
\begin{center}
\resizebox{4in}{!}{\includegraphics{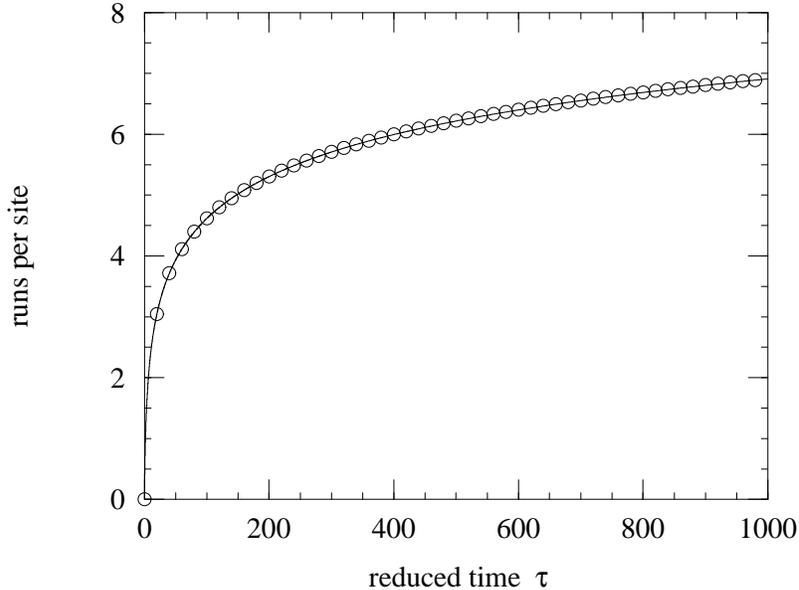}}
\end{center}
\caption{Total number of ``runs'' per site as a function of
  reduced time.  Points are simulation data for $10^9$ time-steps with
  $N=10^6$, and the solid curve is the analytic result, Eq.~\eref{mean}.}
\label{runs}
\end{figure}

The total number of times that the fitness of a given agent gets changed
during the course of the system's life also shows interesting behaviour.
In theory, the number of runs in which each agent takes part, which is
distributed according to the binomial distribution, Eq.~\eref{binomial},
becomes increasingly sharply peaked about its mean as $\tau_f\to\infty$.
Variation in the number of times a given agent is hit will then be a result
of variation only in which particular set of runs affect each agent.  Such
sets are chosen independently at random from the
distribution~\eref{solnpn}.  Since this is a non-normalizable distribution,
the sum representing the total number of hits violates the central limit
theorem---its distribution will have a central portion which is
approximately normally distributed, but there will be a tail of high values
that, like the underlying distribution, goes as $1/n$, being the result of
rare but statistically significant outliers sampled from the tail of the
original distribution.

In practice, however, since the number of runs in which an agent takes part
increases only as the logarithm of~$\tau_f$, Eq.~\eref{mean}, its value is
small for all practical lifetimes of the system and hence shows substantial
statistical fluctuation about its mean.  In typical situations therefore,
the variation in total number of hits an agent receives is dominated by
these fluctuations, giving rise to a distribution which has an exponential
tail.  Only for exponentially long runs $\log(\tau_f+1)\gg1$ will the
power-law behaviour be seen.  Long though they are, none of the simulations
we have performed fall into this category.

\begin{figure}[t]
\begin{center}
\resizebox{4in}{!}{\includegraphics{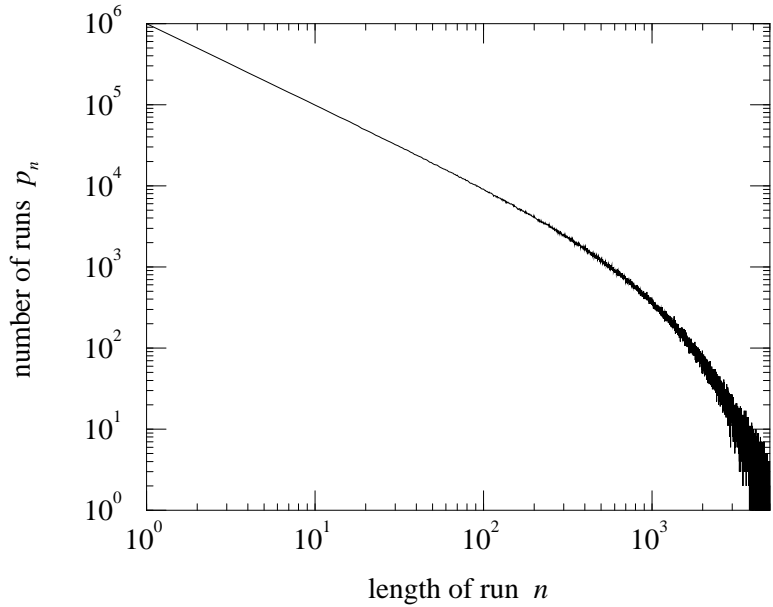}}
\end{center}
\caption{Histogram of the lengths of runs in a simulation of the model for
  $10^9$ time-steps with $N=10^6$.}
\label{pn}
\end{figure}

\section{Numerical results}
Yee's model is straightforward to simulate on a computer, but the naive
method of simulation, in which one searches through all agents on each
time-step to find the least fit, is slow---it takes time $\O(N)$ to perform
the search and hence the entire simulation takes time $\O(t_f N)=\O(\tau_f
N^2)$ to run for reduced time~$\tau_f$.  A better approach is to store the
agents' fitnesses in a binary heap (a partially ordered binary tree with
its smallest value at its root).  This data structure allows us to find the
agent with the lowest fitness in time $\O(1)$, and add or remove agents in
time~$\O(\log N)$, improving the running time of the algorithm to
$\O(\tau_f N\log N)$, which is fast enough for the simulation of quite
large systems.  For a description of the working and implementation of a
binary heap, see for example Refs.~\onlinecite{Knuth73}
and~\onlinecite{AMO93}.

In Fig.~\ref{xt} we show simulation results for $x(\tau)$ as a function of
$\tau$, along with the exact solution~\eref{solnx}, and as the figure
shows, agreement between simulation and exact results is excellent.  In
Fig.~\ref{runs} we show the total number of runs per site $R/N$ as a
function of $\tau$ during a simulation, along with the expected mean
result, Eq.~\eref{mean}, and again the agreement is good.  In Fig.~\ref{pn}
we show on logarithmic scales a histogram of the distribution of the
lengths $n$ of the intervals between evolutionary events.  A straight-line
fit to the data for smaller values of $n$ indicates that the distribution
is following a power-law with exponent $-1.02\pm0.01$, in good agreement
with the predicted exponent of~$-1$, Eq.~\eref{solnpn}.  For larger values
of $n$, deviation from the straight-line form is clearly visible in the
region where $n$ is of the order of $\tau_f$ or greater, where
$\tau_f=1000$ in this case.

\section{Conclusions}
We have analysed the simple evolutionary process proposed by
Yee~\cite{Yee01} in which the agent with the lowest fitness in a large
population is repeatedly removed and replaced with another having fitness
chosen uniformly at random in a given interval.  We have shown that many
properties of the dynamics of this process can be calculated analytically.
Using a numerical method employing a binary heap data structure,
simulations of large realizations of the model are possible in reasonable
time, and we have presented results from such simulations which confirm the
analytic results.

\section*{Acknowledgements}
The authors would like to thank Amnon Aharony for useful conversations.
This work was supported in part by the National Science Foundation.

\end{document}